\documentclass[osajnl,twocolumn,showpacs,superscriptaddress,10pt]{revtex4-1}
\usepackage{mathtools}
\usepackage{amssymb}
\usepackage{amsmath}
\usepackage{amsfonts}
\usepackage{graphicx}
\usepackage{color}
\usepackage{braket}

\begin{document}

\title{Analysis of optical differential transmission signals from co-propagating fields in a lambda system medium}

\author{J.~P.~de Jong}\thanks{Correspondence should be addressed to: jakko.de.jong@rug.nl}
\author{A.~R.~Onur}
\affiliation{Zernike Institute for Advanced Materials,
University of Groningen, NL-9747AG Groningen, The
Netherlands}
\author{D.~Reuter}\thanks{Now at Department of Physics, University Paderborn, Warburger Stra{\ss}e 100, 30098 Paderborn, Germany.}
\author{A.~D.~Wieck}
\affiliation{Angewandte Festk\"orperphysik, Ruhr-Universit\"at Bochum, D-44780 Bochum, Germany}
\author{C.~H.~van~der~Wal}
\affiliation{Zernike Institute for Advanced Materials,
University of Groningen, NL-9747AG Groningen, The
Netherlands}

\date{\today}

\begin{abstract}
We analyze theoretically and experimentally how nonlinear differential-transmission spectroscopy of a lambda system medium can provide quantitative understanding of the optical dipole moments and transition energies. We focus on the situation where two optical fields spatially overlap and co-propagate to a single detector. Nonlinear interactions give cross-modulation between a modulated and non-modulated laser field, yielding differential transmission signals. Our analysis shows how this approach can be used to enhance the visibility of relatively weak transitions, and how particular choices in the experimental design minimize systematic errors and the sensitivity to changes in laser field intensities. Experimentally, we demonstrate the relevance of our analysis with spectroscopy on the donor-bound exciton system of silicon donors in GaAs, where the transitions from the two bound-electron spin states to a bound-exciton state form a lambda system. Our approach is, however, of generic value for many spectroscopy experiments on solid-state systems in small cryogenic measurement volumes where \emph{in-situ} frequency or polarization filtering of control and signal fields is often challenging.
\end{abstract}


\maketitle

\section{\label{sec:intro}Introduction}

Laser spectroscopy is a main research tool to study electronic states of materials, across research fields ranging from atomic physics to cellular biology. In modern optics, nonlinear behavior of a medium is of primary interest, often studied with multiple laser fields which spatially overlap. In such cases, the presence of a strong control laser field makes it challenging to extract the signal from a weak probe field, in particular when further constraints enforce co-propagation of these fields. In practice one cannot always rely on spectral or polarization filtering (for example in cryogenic measurement volumes \cite{Hogele2008,Sladkov2010,Sladkov2011}) and in these cases low-frequency intensity modulation combined with lock-in detection techniques provides a means to separate a signal from co-propagating fields. Such modulation techniques, otherwise known as \emph{optomodulation spectroscopy} \cite{Bonadeo}, also improve the signal to noise ratio \cite{Demtroder}, as signals become less sensitive to external light and $1/f$ noise sources.

With this modulation technique applied to a nonlinear medium, the susceptibilities as seen by the control and probe fields become time dependent. This yields cross-modulation, where modulation of the incident intensity of one field influences the transmitted intensity of another field, giving rise to so-called differential transmission signals. Notably, for this paper the notion of differential transmission refers to this slow modulation of a susceptibility with continuous-wave lasers, effectively comparing two different steady-state conditions.
The modulation timescale is orders of magnitude slower than the intrinsic electronic dynamics of the medium (that can be directly studied in pump-probe experiments with ultrafast pulsed lasers \cite{Sosnowski,Woggon,OLeary,Phelps}, sometimes called \emph{transient differential spectroscopy}).

Low-frequency intensity modulation has already been applied in a variety of optical experiments with differential signals from transmission, reflection or luminescence. Textbooks on nonlinear spectroscopy qualitatively discuss the benefits of studying crosstalk signals from modulation techniques in saturation spectroscopy \cite{Demtroder,AllenEberly,Boyd}. Generically, the advantages lie in enhancing the visibility of weak transitions, and it was applied in experimental work on quantum wells\cite{OLeary}, donor centers in semiconductors \cite{Sladkov2010,Phelps}, quantum dots \cite{Warburton,Bonadeo,Stievater,Silverman}, and other atomic-like systems \cite{Acosta}.
However, a detailed analysis of the differential signals is mostly not presented, despite the fact that this is essential for optimal design of these experiments, interpretation of the results, and insight into systematic errors that may occur.

We present here an analysis of the signals from differential transmission spectroscopy (DTS) on a medium with three-level lambda systems, based on both model calculations and experimental results. We show how DTS can dramatically enhance the spectroscopic visibility of weakly absorbing transitions, and how this enhancement depends on the involved optical dipole moments. We also investigate how the method can provide quantitative information about these dipole moments (which are key ingredients if one also wishes to unravel the combined DTS signal from two co-propagating fields into the separate transmission components). Further, our analysis shows how in two-laser experiments (where one laser is fixed and one is used for spectroscopic scanning) choosing which laser is modulated makes a key difference for being robust against unintentional intensity variations in the applied fields. Similarly, we show how the spectroscopic lines of a transition can show an apparent frequency shift. We show quantified results for this shift and describe the best approach for minimizing this systematic error.

The relevance of our analysis is demonstrated with spectroscopy experiments on donor-bound exciton centers in Si-doped insulating gallium arsenide. While our experiments and part of our calculations thus focus on one particular material system and experimental arrangement, the analysis and methods that we present here are of generic value for spectroscopy experiments on media with multi-level systems.

The paper is organized as follows. Section~\ref{sec:simu} presents simulations of DTS on a three-level lambda system medium. We introduce a model to calculate differential signals and describe it both at the macroscopic scale of the measurement setup and in terms of the microscopic electron dynamics. Subsequently, this model is used to analyze the benefits of differential transmission and how possible disadvantages can be minimized. Section~\ref{sec:expSiGaAs} provides experimental results of DTS on silicon donor centers in gallium arsenide, demonstrating the effects described by our model.

\section{\label{sec:simu}Simulations}

\subsection{Model}

\begin{figure}[t!]
\includegraphics[width=86mm]{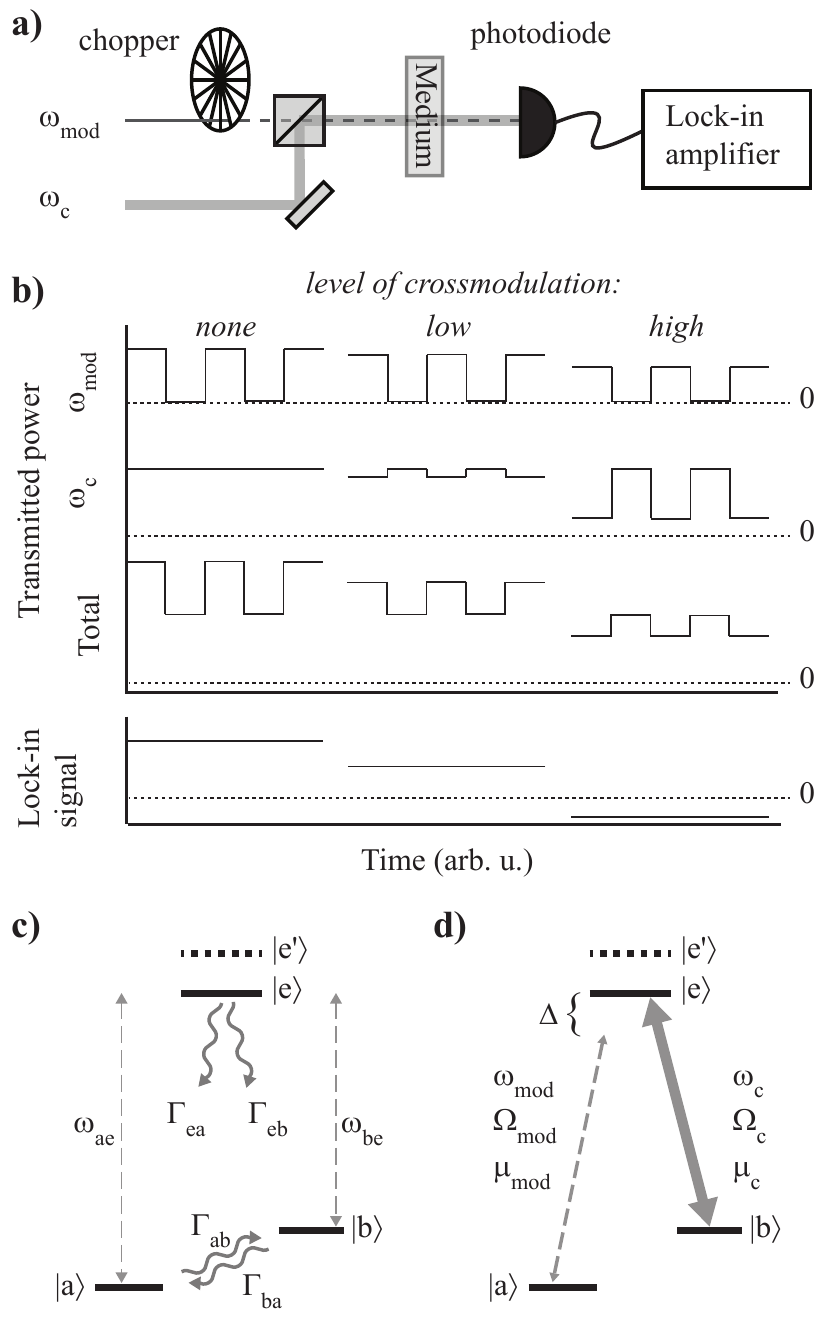}
\caption{ a) Experimental setup for differential transmission spectroscopy. Two overlapping laser fields, of which one is modulated by a chopper, co-propagate through the medium. Transmitted light is converted into an electrical signal by a photodiode. A lock-in amplifier filters out the signal component at the modulation frequency. b) Schematic illustration of the transmitted power of both laser fields, and the resulting lock-in signal. Signals are plotted for three qualitatively different levels of cross-modulation. c) Four-level $\Lambda$ system with ground states $\ket{a}$ and $\ket{b}$ and excited states $\ket{e}$ and $\ket{e'}$, with optical transitions from both ground states. The transitions of $\ket{a}$ and $\ket{b}$ to $\ket{e}$ have energies $\hbar \omega_{ae}$ and $\hbar \omega_{be}$. Population relaxation rates $\Gamma_{ij}$ are depicted by curly arrows. d) The modulated and constant laser, with frequency $\omega_\text{mod}$ and $\omega_\text{c}$ respectively, couple to the transitions with dipole moment $\mu_\text{mod}$ and $\mu_\text{c}$, resulting in Rabi frequencies $\Omega_\text{mod}$ and $\Omega_\text{c}$. When lasers only address transitions to $\ket{e}$ the system behaves as a three-level system. One laser is coupled resonantly, while the other (in this figure the modulated laser) is scanned over resonance by changing the detuning $\Delta$. }
\label{fig:fig1}
\end{figure}

We model a setup where two continuous-wave laser fields co-propagate through a medium, as shown in Fig.~\ref{fig:fig1}(a). One field, depicted by its angular frequency $\omega_\text{mod}$, is undergoing on-off modulation at frequency $f_\text{mod}$ by a chopper. The other field, with angular frequency $\omega_\text{c}$, is not modulated and hence has a constant intensity. The total transmitted power is converted into an electrical signal by a linearly behaving photodiode after the sample. A lock-in amplifier isolates the electrical component at frequency $f_\text{mod}$ from all other components.

The modulation is orders of magnitude slower than the electronic dynamics in the medium and is assumed to have a square on-off envelop. This leads to two steady-state situations with transmittance $T^\text{on}_{\omega_\text{mod}}$ for the modulated field and transmittances $T^\text{on}_{\omega_\text{c}}$ and $T^\text{off}_{\omega_\text{c}}$ for the constant field. If the probed medium has nonlinear components in the susceptibility, \textit{i.e.} the susceptibility depends on the presence of laser fields, $T^\text{on}_{\omega_\text{c}}$ and $T^\text{off}_{\omega_\text{c}}$ are in general not equal. Consequently, the transmittance of the constant field is time-dependent with frequency $f_\text{mod}$ in the form of amplitude modulation of its transmission. This transfer of amplitude modulation from one field to another via the susceptibility will be called cross-modulation, and the contribution of the constant field to the transmission at frequency $f_\text{mod}$ will be called differential transmission. The frequency component at $f_\text{mod}$ of the total transmitted power thus consists of two parts and is given by
\begin{equation}
P_{f_\text{mod}} = \underbrace{P_\text{mod}~T^\text{on}_{\omega_\text{mod}}}_\text{normal transmission} + \underbrace{P_\text{c} (T^{\text{on}}_{\omega_\text{c}}-T^\text{off}_{\omega_\text{c}})}_\text{differential transmission}
\label{eq:totalPower}
\end{equation}
where $P_\text{mod}$ ($P_\text{c}$) is the power of the modulated (constant) field,
and where $P_\text{mod}$ and $P_\text{c}$ are incident on the medium while $P_{f_\text{mod}}$ is measured after the medium. We will focus on scenarios with $P_\text{mod} < P_\text{c}$ to ensure a prominent role for the differential transmission. Figure~\ref{fig:fig1}(b) illustrates the transmitted power for the individual fields and the total transmitted power, for qualitatively different levels of cross-modulation. If cross-modulation is absent, the contribution of differential transmission in Eq.~(\ref{eq:totalPower}) is zero, and the lock-in signal consists purely of the normal transmission of the modulated field. For increasing levels of cross-modulation, the contribution of the differential transmission to the lock-in signal increases. When the differential transmission is larger then the normal transmission, and of opposite sign (which is the case for lambda systems), the total transmitted power shows a 180$^\circ$ phase shift. This shift manifests itself in the lock-in signal either as a negative signal or a 180$^\circ$ phase shift, depending on lock-in operation settings (for examples see Fig.~\ref{fig:fig5} and Fig.~\ref{fig:fig4}(c), respectively).

The medium of interest consists of lambda systems, as shown in Fig.~\ref{fig:fig1}(c). Two ground states $\ket{a}$ and $\ket{b}$ have optical transitions to common excited states $\ket{e}$, $\ket{e'}$, \textit{etc.} There is no optical transition between $\ket{a}$ and $\ket{b}$. Relaxation parameters $\Gamma_{ij}$ describe both spontaneous emission rates from the excited states to the ground states and thermalization of population in the ground states. Furthermore, all states except $\ket{a}$ undergo pure dephasing $\gamma_b$, $\gamma_e$, \textit{etc}. For the sake of simplicity, the simulations will be restricted to three-level lambda systems, without loss of validity. The experimental results in Sec.~\ref{sec:expSiGaAs}, however, show multiple excited states.

Differential transmission spectroscopy is modeled with two laser fields coupled to the optical transitions with transition dipole moments $\mu_\text{mod}$ and $\mu_\text{c}$, see Fig.~\ref{fig:fig1}(d). One field is held resonant with its transition frequency, while the other is scanned over the resonance by changing its detuning $\Delta$. It is assumed that each laser couples only to one transition. The lasers drive transitions between the levels at Rabi frequencies $\Omega_\text{mod}$ and $\Omega_\text{c}$. The population in $\ket{e}$ spontaneously decays to both ground states, with relaxation rates $\Gamma_{ij}$. Hence, a field present at transition $\omega_{ae}$ ($\omega_{be}$) will effectively pump population to state $\ket{b}$ ($\ket{a}$), increasing the absorption coefficient for the field at transition $\omega_{be}$ ($\omega_{ae}$). In Eq.~(\ref{eq:totalPower}) this results in the contribution $(T^{\text{on}}_{\omega_\text{c}}-T^\text{off}_{\omega_\text{c}})$, which is always negative for lambda systems. For the present analysis, we neglect effects related to coherent population trapping that can take place in a narrow spectral window around two-photon resonance when two lasers drive a lambda system \cite{Fleischhauer2005,Sladkov2010}.
The amount of population pumped from one ground state of the lambda system to the other depends on the ratio of relaxation coefficients $\Gamma_{ea}$ and $\Gamma_{eb}$. The relaxation coefficient of a transition is related to its electric dipole moment by
\begin{equation}
\Gamma_{ij} = \frac{2 n \omega_{ij}^3 \mu_{ij}^2}{3\epsilon_0 h c^3}
\end{equation}
where $n$ is the bulk refractive index. 
The branching ratio $\Gamma_{ea} /\Gamma_{eb}$  is given by
\begin{equation}
\frac{\Gamma_{ea}}{\Gamma_{eb}} = \frac{\omega_{ea}^3 \mu_\text{mod}^2}{\omega_{eb}^3 \mu_\text{c}^2} \approx \left( \frac{\mu_\text{mod}}{\mu_\text{c}}\right)^2 \, , \quad \text{for} \; \omega_{ae} \approx \omega_{be}
\end{equation}
The ratio $\mu_\text{mod}/\mu_\text{c}$, which we will call \emph{relative dipole moment}, is the main parameter that determines the amount of cross-modulation and differential transmission (Fig.~\ref{fig:fig2}).

The transmittances are determined by the imaginary part of the susceptibility by
\begin{equation}
T(\omega) = \text{exp}(-\frac{z\omega}{c} \text{Im}[\chi(\omega)])
\end{equation}
where $z$ is the sample thickness and $c$ is the speed of light in vacuum. For each field $\chi(\omega)$ is calculated by
\begin{equation}
\chi(\omega) = \frac{2N\mu_{ij}^2\sigma_{ij}(\omega)}{\epsilon_0 \hbar \Omega}
\end{equation}
where $N$ is the number density of lambda systems in the material. The slowly oscillating part of the transition's coherence $\sigma_{ij}$ is obtained from the steady-state solution of the master equation for the density operator
\begin{equation}
\frac{d\hat{\rho}}{dt} = - \frac{i}{\hbar} [\hat{H},\hat{\rho}] + \hat{\mathcal{L}}(\hat{\rho})
\end{equation}
where $\hat{H}$ is the Hamiltonian of the laser-driven system in Fig.~\ref{fig:fig1}(c,d) (using a standard approach \cite{Fleischhauer2005}). The Lindblad operator $\hat{\mathcal{L}}$ includes all relaxation and decoherence rates. We solve the master equation numerically, applying the rotating wave approximation \cite{Fleischhauer2005}.

\subsection{Results}

\begin{table}[h]
\centering
\caption{Simulation parameters}
\label{tab:params}
\begin{tabular}{c l p{1cm} c l}
\hline
$\Gamma_{ea}+\Gamma_{eb}$ & 1 GHz & & $\Gamma_{ba}$ & 0.37 MHz \\
$\gamma_e$ & 5 GHz & & $z$ & 10 $\mu$m \\
$\gamma_b$ & 5 GHz & & $N$ & $10^{-13}~\text{cm}^{-3}$ \\
\hline
\end{tabular}
\end{table}

\begin{figure}[t!]
\includegraphics[width=86mm]{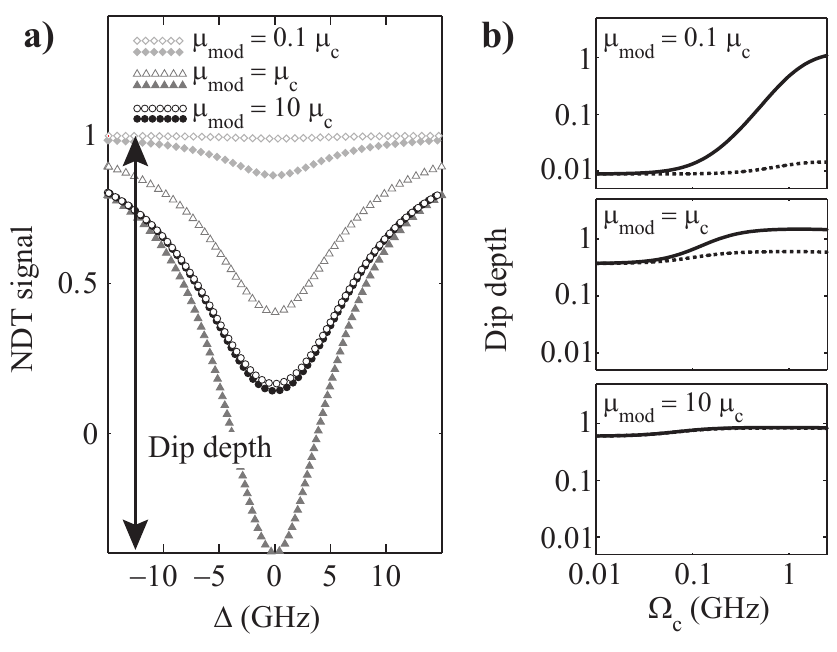}
\caption{Simulation of normalized differential transmission (NDT) for three different relative dipole moments $\mu_\text{mod} / \mu_\text{c}$. a) NDT while scanning the modulated beam over the resonance, with $\Omega_\text{mod} = 10~\text{MHz}$ and $\Omega_\text{c} = 0.5~\text{GHz}$. The traces show the total signal (solid symbols) as seen by the lock-in, which consists of normal transmission of the modulated field and differential transmission of the constant field due to cross-modulation. The open symbols show the contribution of the normal transmission. b) Enhanced transition visibility as a function of Rabi frequency of the constant laser, with $\Omega_\text{mod} = 10~\text{MHz}$. The full lines show the dip depth of the total signal, while the dashed lines show only the normal transmission of the modulated field.}
\label{fig:fig2}
\end{figure}

The simulations presented below were obtained with parameters that resemble our experimental system, see Table \ref{tab:params}. However, our model yields qualitatively similar results for a large range of parameters.

Figure~\ref{fig:fig2}(a) shows DTS traces for three different values for the relative dipole moment. The absolute values of the dipole moments were varied in a manner that keeps the total relaxation rate $\Gamma_{ea}+\Gamma_{eb}$ from the excited state the same for the three cases. The solid lines present results of Eq.~(\ref{eq:totalPower}) normalized to $P_\text{mod}$, and represent the total signal as seen by the lock-in. The dashed lines present (on the same scale) the normal transmission of the modulated laser, such that the difference with the solid lines signifies the magnitude of the differential transmission part. The Rabi frequencies are $\Omega_\text{mod}=10$~MHz for the modulated laser and $\Omega_\text{c}=0.5$~GHz for the constant laser, for all three cases. A normalized measure of the transition visibility, which we will call \emph{dip depth}, can be assigned to each scan as shown for the deepest scan. Figure~\ref{fig:fig2}(b) shows the dip depth as a function of $\Omega_\text{c}$, again comparing the total lock-in signal with the normal transmission part.

The visibility of the transition is strongly enhanced by cross-modulation for lambda systems with a small relative dipole moment (shown for $\mu_\text{mod}/\mu_\text{c} = 0.1$ in the top panel of Fig.~\ref{fig:fig2} (b)). At high $\Omega_\text{c}$, the differential transmission is two orders of magnitude larger than the normal transmission. For systems with a large relative dipole moment (bottom panel Fig.~\ref{fig:fig2}(b)), differential transmission is negligible, since the amount of cross-modulation is low. The absorption of the modulated laser is sufficient to have a good transition visibility at both low and high  $\Omega_\text{c}$. Here the presence of a strong non-modulated laser only slightly enhances the absorption of the modulated beam and hence the dip depth. In the intermediate case where $\mu_\text{mod}/\mu_\text{c} = 1$, the differential transmission approximately doubles the dip depth (middle panel Fig.~\ref{fig:fig2}(b)). For a range of $\mu_\text{mod}/\mu_\text{c}$ values, strong cross-modulation can increase the dip depth to values greater than 1. This yields a negative DTS signal, as can be seen in Fig.~\ref{fig:fig2}(a) (solid trace for $\mu_\text{mod}/\mu_\text{c} = 1$) and in the top two panels of Fig.~\ref{fig:fig2}(b) at high $\Omega_\text{c}$. Whether this occurs, depends on both the relative and absolute dipole moments. For smaller absolute dipole moments the results are very similar to Fig.~\ref{fig:fig2}, with smaller dip depths.

\begin{figure}[t!]
\includegraphics[width=86mm]{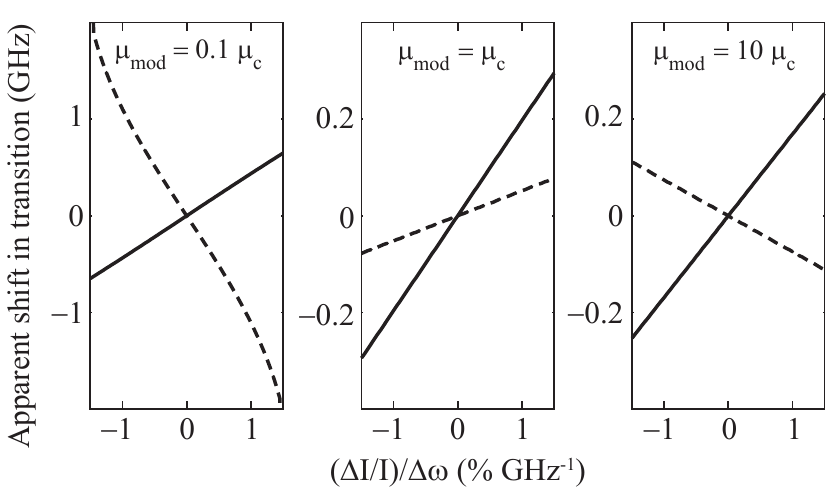}
\caption{Error in the observed transition frequency due to intensity changes during spectroscopy scans for three different relative dipole moments (note different vertical scale for the left panel). The apparent shift is the difference between the lowest point in a spectroscopic dip and the real transition frequency. $(\Delta I / I)/\Delta\omega$ is the percentage change in laser intensity per GHz scan range. Full (dashed) lines show the apparent shift while scanning the constant (modulated) laser. Results for $\Omega_\text{mod} = 10~\text{MHz}$ and $\Omega_\text{c} = 0.5~\text{GHz}$.}
\label{fig:fig3}
\end{figure}

Optical transitions in spectroscopy are observed as peaks or dips against a background signal. When this background has a slope or even additional structure, it is harder to distinguish transitions. Moreover, this causes an offset for the lowest (highest) point of a dip (peak) that should identify the transition frequency. In addition, as we analyze below here, one can have apparent shifts in the transition frequency when the intensity of the scanning laser is not constant while scanning over the resonance.
For this latter effect, the causes can be in the experimental setup (for example changes in laser power during frequency scans) or fundamental to the sample (for example due to Fabry-Perot cavity effects inside the sample, for an experimental example see Fig.~\ref{fig:fig4}(a)).

We modeled the effect of intensity changes on the observed transition frequency, comparing the apparent shift when scanning the modulated or the constant laser. We change the intensity of the scanning laser linearly with frequency. The unit of this intensity change is percentage per GHz scan range, with the intensity at the transition center defined as 100\%. The lowest point in the DTS signal (in simulated traces, in appearance similar to the related experimental results in Fig.~\ref{fig:fig4}(c)) is taken to be the spectroscopic position of the scanned transition. Figure~\ref{fig:fig3} shows the apparent shift as a function of linear power changes during scanning, for different relative dipole moments. Full (dashed) lines show the results for scanning the constant (modulated) laser. Errors in the transition frequency on the order of a GHz can be obtained for realistic changes in intensity over scan range. The error is largest for small relative dipole moments. However, this error can be suppressed a factor of three by scanning the constant laser. On the contrary, larger relative dipole moments show an enhanced spectroscopic error when the constant laser is scanned. Scanning the constant laser always results in a transition shift with the same sign as the intensity change (solid lines in the three panels). The sign of the error when scanning the modulated laser is less straightforward and depends on the competition between opposite shifts in the normal and differential transmission parts of Eq.~(\ref{eq:totalPower}).

To derive quantitative information on the relative and absolute strength of the optical transition dipole moments from such DTS data, a single DTS trace does not provide sufficient information. Instead, one needs to take data for a range of $\Omega_\text{mod}$ and $\Omega_\text{c}$ values, and fit the observed trends in dip depth to traces as in Fig.~\ref{fig:fig2}. If not known from independent measurements, the fitting should also yield parameters for the decay and dephasing parameters, and the optical Rabi frequencies. The nonlinear behavior of the overall system allows for such a multi-parameter fitting analysis. Notably, this approach is thus also needed if one wishes to unravel the combined DTS signal from two co-propagating fields into the separate transmission components.

\section{\label{sec:expSiGaAs}Experiments on S\lowercase{i} donors in \lowercase{i}-G\lowercase{a}A\lowercase{s}}

\subsection{Material and methods}
We used DTS as a spectroscopic technique to find the optical transitions in lambda systems provided by silicon donors in epitaxial insulating gallium arsenide. We studied epitaxially grown samples of 10~$\mu$m thickness and with a donor concentration of $\sim$$3 \times 10^{13}~{\rm cm^{-3}}$. The samples were studied at a temperature of $4.2$~K and an applied magnetic field $B$ of several Tesla. At such low donor concentration and temperature the donor centers are not ionized, thus providing an ensemble of localized electrons (so-called $D^0$ systems). These electrons have selectively-addressable optical transitions from the two Zeeman-split spin $S=\frac12$ levels of the $D^0$ ground state to a donor-bound exciton complex ($D^0 X$ system). The $D^0 X$ system consists of two electrons and one hole bound at the donor side. This system has several energy levels in a frequency window of tens of GHz due to interactions between the particles, quantum confinement around the donor site and Zeeman shifts from the spin of the three particles \cite{Karasyuk1994,Sladkov2010}. In our study the propagation of optical fields was in Voigt geometry. Two co-propagating laser fields with linear polarization were guided to the sample through a single mode fiber, and we had a detector directly behind the sample for recording DTS signals (for further details see Refs.~\onlinecite{Sladkov2010,Sladkov2011}). 

For applying the DTS technique and getting the results in Figs.~\ref{fig:fig5} and \ref{fig:fig4}, one laser was fixed at the optical transition in the spectrum from the $D^0$ spin-down state to the lowest level $\ket{e}$ of the $D^0 X$ complex (see Fig.~\ref{fig:fig1}(c.d)), while the other laser was scanned to probe the transitions from the $D^0$ spin-up state to the sequence of levels $\ket{e}$, $\ket{e'}$ \textit{etc}. For the examples presented here, the polarization of the fixed laser was parallel to the magnetic field, and that of the scanning laser either parallel or orthogonal \cite{Sladkov2010}. Strictly speaking, this two-laser driving only truly addressed a three-level lambda system when probing level $\ket{e}$. However, for the subsequent levels $\ket{e'},...$ the two-laser approach counter-acts optical pumping into one of the $D^0$ spin states, and for the investigation of how this yields cross-modulation signals it can for each transition be analyzed using our three-level modeling.

\subsection{Results}
\begin{figure}[t!]
	\includegraphics[width=86mm]{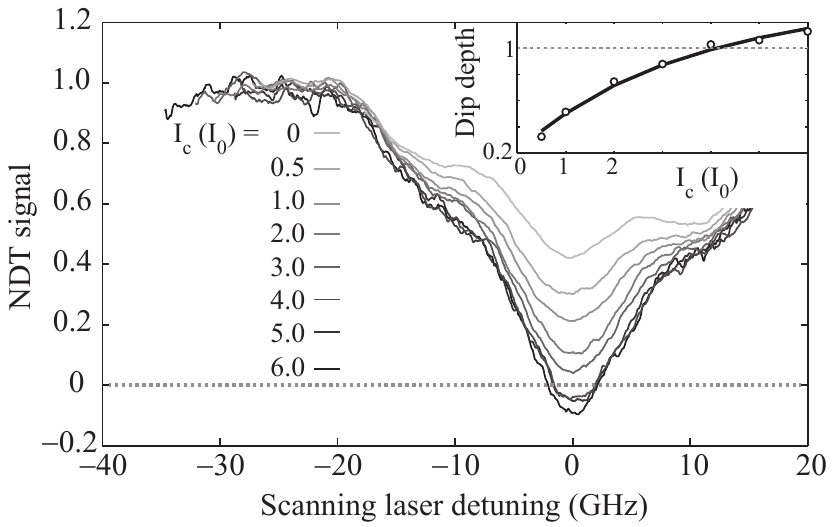}
	\caption{Normalized differential transmission (NDT) on a lambda system of the donor-bound exciton complex in GaAs:Si at $B = 3.94~\text{T}$. The constant laser is kept resonant, while the modulated laser is scanned over the other transitions of the lambda system. Increasing the intensity of the constant laser $I_\text{c}$ enhances the transition visibility ($I_0 = 0.7~\text{mW}~\text{cm}^{-2}$). For large $I_\text{c}$, the dip depth becomes greater than one. We fit the dip depth of the experimental traces with our model, using the dipole moments and laser spotsize as fit parameters. The inset shows the experimentally found dip depths (circular symbols) and the resulting fit (full line), with $\mu_\text{mod} = 1.38~e\cdot\text{nm}$ and $\mu_\text{c} = 2.15~e\cdot\text{nm}$.}
	\label{fig:fig5}
\end{figure}

Figure~\ref{fig:fig5} presents experimental DTS traces, taken with the constant laser fixed as described in the previous paragraph and the modulated laser scanning over the transition from the $D^0$ spin-up state to the third level in the $D^0 X$ complex. Resonance with this $D^0$--$D^0 X$ transition appears as a dip in transmission. The NDT signal is presented as a function of the detuning from resonance, and is normalized at the value at off-resonance detuning. Between traces, the intensity of the constant laser is increased. Since the spotsize of the lasers was uncertain in this specific experiment, $I_c$ is provided as a multiple of $I_0$ (with the value $0.7~\text{mW}~\text{cm}^{-2}$, see below). The transitions visibility increases with $I_c$, and the NDT signal drops below zero for the highest intensities. Strong variations in the laser intensity show as additional wiggles on the NDT signal. These variations are caused by a weak Fabry-Perot effect inside the sample, experienced by the scanning laser. To find the dip depth as a function of $I_c$, we subtract the background signal, as observed with the constant laser blocked ($I_c = 0$), from the other traces. For the present case this yields a good approximation for differential transmission contribution to the full DTS signal. The evolution of the traces with $I_c$ indicates a relative dipole moment $\mu_\text{mod}/\mu_\text{c} \lesssim 1$. This makes the differential transmission part the dominant term in Eq.~\ref{eq:totalPower} (see also Fig.~\ref{fig:fig2}(b)), justifying this approach. The inset of Fig.~\ref{fig:fig5} shows the resulting dip depths (circles) and a fit from our model (line, a trace similar to the solid lines in Fig.~\ref{fig:fig2}(b)). The corresponding fit parameters are $\mu_\text{mod} = 1.38~e\cdot\text{nm}$, $\mu_\text{c} = 2.15~e\cdot\text{nm}$ and $I_0 = 0.7~\text{mW}~\text{cm}^{-2}$.

\begin{figure}[t!]
\includegraphics[width=86mm]{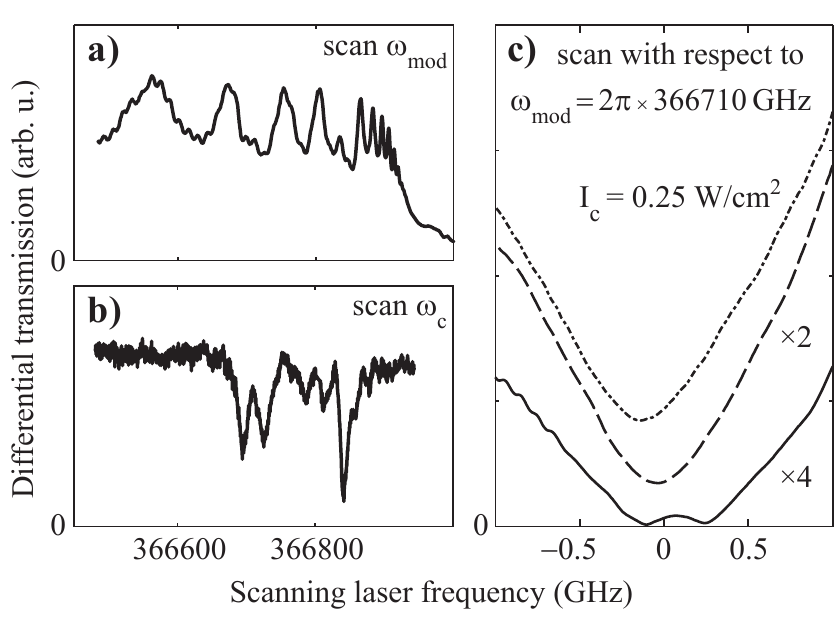}
\caption{Experimental data of differential spectroscopy on donor-bound electrons in GaAs at $B = 6.41~\text{T}$, produced by scanning a) the modulated laser or b) the constant laser. In a) the spectroscopic features are imposed on a strongly varying background (in this specific case due to a weak Fabry-Perot interference in the sample). The approach used for b) is less influenced by Fabry-Perot interference. Here the various spectroscopic lines (corresponding to a set of donor-bound exciton states) appear on a flat background. c) Close-up of the spectroscopic line at $\sim$366710~GHz, for different constant laser intensities, obtained with the approach of panel a). Both the visibility and the apparent spectroscopic position of the transition are changing with laser intensity. The signal crosses zero for the highest intensity (here plotted as a positive signal for which it was noted that the lock-in signal had a 180 degree phase change).}
\label{fig:fig4}
\end{figure}

The influence of intensity variations in the scanning laser on the DTS signal can be minimized by scanning the constant laser, as illustrated by comparing Fig.~\ref{fig:fig4}(a) and Fig.~\ref{fig:fig4}(b). These panels show similar DTS scans, but differ in which laser was scanned. In Fig.~\ref{fig:fig4}(a), the modulated laser is scanned. Its intensity undergoes significant variations due to the weak Fabry-Perot interference (which is chirped due to the very strong free-exciton absorption line at $\sim$367000~GHz). Since the modulated laser is scanned, these variations contribute to the normal transmission, and via cross-modulation also to the differential transmission near a resonance. This severely deteriorates the visibility of the transitions. Conversely, in Fig.~\ref{fig:fig4}(b) the constant laser is scanned while the modulated laser is fixed at resonance. The constant laser scans over the same Fabry-Perot pattern as in Fig. \ref{fig:fig4}(a), but the resulting intensity variations do not contribute to the DTS signal when the scanning laser is off-resonance. Transitions now appear on a flat background. Notably, one can clearly recognize at least six $D^0$--$D^0 X$ resonance lines in the data. This technique hence provides clear high resolution spectroscopy data, not cluttered with the underlying Fabry-Perot pattern.

Figure~\ref{fig:fig4}(c) shows traces taken with the approach as for Fig.~\ref{fig:fig4}(a), zooming in on one particular transition. The intensity of the constant laser is varied between traces. An apparent shift in the transition frequency of hundreds of MHz can be observed with increasing intensity. Additionally, the dip depth exceeds 1 at the highest value for $I_c$ (here shown as a lock-in phase flip). We ran simulations for the apparent transition shift as in Fig. \ref{fig:fig3}, obtaining a realistic value for $(\Delta I / I)/\Delta\omega$ from the Fabry-Perot pattern in Fig. \ref{fig:fig4}(a). The simulated shifts agreed with the experimental shift within a factor two.

\section{Conclusion}
We have analyzed how low-frequency amplitude modulation in two-laser spectroscopy yields differential signals for lock-in detection with a single photodiode. We applied this to DTS studies of a medium with lambda systems. Cross-modulation provides a very useful tool for enhancing the visibility of transitions that are otherwise difficult to observe. We quantified this enhancement and showed how the resulting spectroscopy provides information about the underlying dipole moments. Choosing which laser to scan can improve signal disturbance from laser-field intensity changes, and minimize the spectroscopic error due to these effects. Experimental DTS results on GaAs:Si show the relevance and validity of our model. We envision that our modeling can be used for spectroscopy on a wider range of systems with multiple levels in the ground and excited state.

\vspace{3mm}

We thank D.~O'Shea, A.~U.~Chaubal, and R.~Lous for valuable discussions and experimental aid, and acknowledge financial support from the Dutch FOM and NWO, a European ERC Starting Grant, the Research school Ruhr-Universit\"{a}t Bochum, and the German programs BMBF Q.com-H 16KIS0109, Mercur Pr-2013-0001, and the DFH/UFA CDFA-05-06.



\begin{thebibliography}{99}

\bibitem{Hogele2008} A.~H\"ogele, S.~Seidl, M.~Kroner, K.~Karrai, C.~Schulhauser, O.~Sqalli,
J.~Scrimgeour, and R.~J.~Warburton, "Fiber-based confocal microscope for cryogenic spectroscopy," Rev. Sci. Instr. \textbf{79}, 023709 (2008).

\bibitem{Sladkov2010} M.~Sladkov, A.~U.~Chaubal, M.~P.~Bakker, A.~R.~Onur, D.~Reuter, A.~D.~Wieck, C.~H.~van~der~Wal, "Electromagnetically induced transparency with an ensemble of donor-bound electron spins in a semiconductor," Phys. Rev. B \textbf{82}, 121308(R) (2010).

\bibitem{Sladkov2011} M.~Sladkov, M.~P.~Bakker, A.~U.~Chaubal, D.~Reuter, A.~D.~Wieck, C.~H.~van~der~Wal, "Polarization-preserving confocal microscope for optical experiments in a
dilution refrigerator with high magnetic field," Rev. Sci. Instr. \textbf{82}, 043105 (2011).

\bibitem{Bonadeo} N.~H.~Bonadeo, A.~S.~Lenihan, Gang Chen, J.~R.~Guest, D.~G.Steel, D.~Gammon, D.~S.~Katzer and D.~Park, "Single quantum dot states measured by optical modulation spectroscopy," Appl. Phys. Lett. \textbf{75}, 2933 (1999).

\bibitem{Demtroder} W.~Demtr\"oder, \emph{Laser Spectroscopy 2nd ed.} (Springer, 1998).

\bibitem{Sosnowski} T.~S.~Sosnowski, T.~B.~Norris, H.~Jiang, J.~Singh, K.~Kamath, P.~Bhattacharya, "Rapid carrier relaxation in In$_{0.4}$Ga$_{0.6}$As/GaAs quantum dots characterized by differential
transmission spectroscopy," 
Phys. Rev. B \textbf{57}, R9423(R) (1998).

\bibitem{Woggon} U.~Woggon, H.~Giessen, F.~Gindele, O.~Wind, B.~Fluegel and N.~Peyghambarian, "Ultrafast energy relaxation in quantum dots," Phys. Rev. B \textbf{54}, 17681-17690 (1996).

\bibitem{OLeary} S.~O'Leary, H.~Wang, J.~P.~Prineas, "Coherent Zeeman resonance from electron spin
coherence in a mixed-type GaAs/AlAs quantum well," Optics Letters \textbf{32}, 569-571 (2007).

\bibitem{Phelps} C.~Phelps, S.~O'Leary, J.~Prineas, H.~Wang, "Coherent spin dynamics of donor bound electrons in GaAs," Phys. Rev. B \textbf{84}, 085205 (2011).

\bibitem{AllenEberly} L.~Allen, J.~H.~Eberly, \emph{Optical Resonance and Two-Level Atoms 2nd ed.} (Dover Publications, 1987).

\bibitem{Boyd} R.~W.~Boyd, \emph{Non-linear Optics 3rd ed.} (Academic Press, 2008).

\bibitem{Warburton} K.~Karrai, R.~J.~Warburton, "Optical transmission and reflection spectroscopy of
single quantum dots," \textbf{33}, 311-337 (2003).

\bibitem{Stievater} T.~H.~Stievater, Xiaoqin~Li, T.~Cubel, D.~G.Steel, D.~Gammon, D.~S.~Katzer and D.~Park, "Measurement of relaxation between polarization eigenstates in single quantum dots," Appl. Phys. Lett. \textbf{81}, 4251-4253 (2002).

\bibitem{Silverman} K.~L.~Silverman, R.~P.~Mirin, S.~T.~Cundiff, "Lateral coupling of In$_x$Ga$_{1-x}$As/GaAs quantum dots investigated using differential transmission spectroscopy," Phys. Rev. B \textbf{70}, 205310 (2004).

\bibitem{Acosta} V.~M.~Acosta, K.~Jensen, C.~Santori, D.~Budker, and R.~G.~Beausoleil, "Electromagnetically Induced Transparency in a Diamond Spin Ensemble Enables All-Optical Electromagnetic Field Sensing," Phys. Rev. Lett. \textbf{110}, 213605 (2013).

\bibitem{Fleischhauer2005} M. Fleischhauer, A. Imamoglu, J.P. Marangos, "Electromagnetically induced transparency: Optics in coherent media," Rev. Mod. Phys. \textbf{77}, pp. 633-673 (2005).

\bibitem{Karasyuk1994}
V.~A.~Karasyuk, D.~G.~S.~Beckett, M.~K.~Nissen, A.~Villemaire, T.~W.~Steiner, M.~L.~W.~Thewalt, "Fourier-transform magnetophotoluminescence spectroscopy of donor-bound excitons in GaAs," Phys. Rev. B \textbf{49}, 16381, (1994).


\end{thebibliography}
\end{document}